\documentclass[12pt,twoside,a4paper]{article}
\usepackage{lineno,hyperref,amsmath,float,graphicx}
\usepackage{subfig}
\usepackage[utf8x]{inputenc}

\begin{document}

\begin{center}\LARGE
\textbf{The spiral arms of galaxies}
\end{center}

\begin{center}\small  G. Contopoulos\footnote{gcontop@academyofathens.gr}\\ \textit{Research Center for Astronomy and 
Applied Mathematics \\of the Academy of 
Athens\\ Soranou Efesiou 4, GR-11527 Athens, Greece}
\end{center}

\begin{abstract}
The most important theory of the spiral arms of galaxies is the density wave theory based on the Lin-Shu dispersion relation. However, the density waves move with the group velocity towards the inner Lindblad resonance and tend to disappear. Various mechanisms to replenish the spiral waves have been proposed. Nonlinear effects play an important role near the inner and outer Lindblad resonances and corotation. The orbits supporting the spiral arms are precessing ellipses in normal galaxies that extend up to the 4/1 resonance. On the other hand, in barred galaxies the spiral arms extend along the manifolds of the unstable periodic orbits at the ends of the bar and they are composed of chaotic orbits. However these chaotic orbits can be found analytically.
\end{abstract}

\section{Introduction}

A large proportion of galaxies have spiral arms. But how these spiral arms were formed? Today I will present a historical account of the various theories of spiral arms with emphasis on the density wave theory.

 An obvious theory of spiral arms is  based on the differential rotation of the galaxies. Namely, the rotational velocity $\Omega$ around the center of the galaxy decreases (Fig. 1) as the distance from the center increases. Thus, an initially elongated structure (Fig. 2a) is transformed soon into a spiral, as the regions closer to the center rotate faster, leaving the outer parts behind (Fig. 2b) (trailing spiral areas). These arms are called material arms, because they are composed of the same material always.

\begin{figure}[!h]
\centering
\includegraphics[scale=0.4]{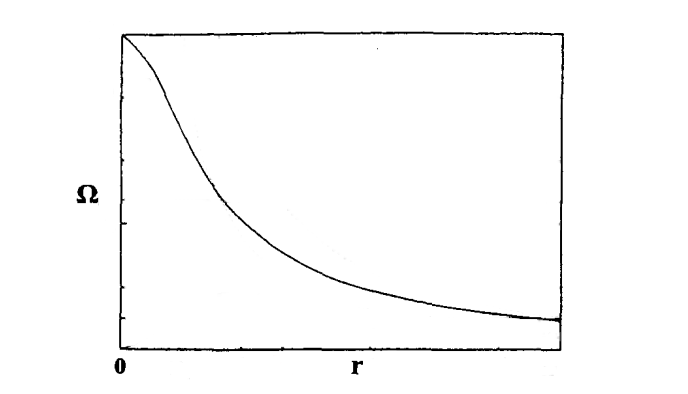}
\caption{Rotational velocities of galaxies.}
\end{figure}

\begin{figure}[!h]
\centering
\includegraphics[scale=0.4]{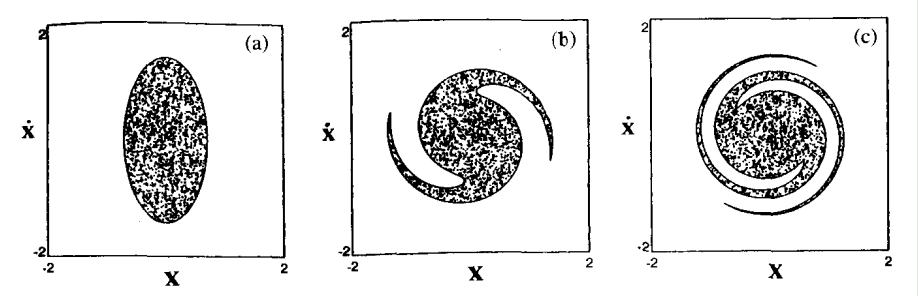}
\caption{The evolution of material spirals}
\end{figure}

However, if we take the data from our Galaxy and similar galaxies, we find that the material arms of the Galaxy were straight lines only 300 million years ago and they should be wrapped tightly after a few rotations (Fig. 2c). Thus the galaxies that are much older should not show the beautiful spiral arms that we observe.

Then some people considered a very different mechanism of generating spiral arms. As B. Lindblad (1926, 1936) has shown, if a galaxy is quite flat (beyond the flatnem of E7 galaxies) the circular motions at its boundary are unstable and matter is ejected outwards in the direction of rotation (leading spiral arms). However, most galaxies are not so flat and the observations show that the spiral arms are trailing, therefore this mechanism is not satisfactory. 

Other mechanisms, like magnetic fields acting on the gas have been shown to be insufficient because the observed magnetic fields are very weak. 

\section{The density wave theory}

B. Lindblad considered the mechanism of density waves (Lindblad 1940, 1941, Lindblad and Langebartel, 1953). Namely the stars move through the spiral arms but stay longer in their neighborhood (Fig. 3). Therefore, the spiral arms are waves. They keep their form, but they are not composed of the same matter for a long time. 
This mechanism was shown to be very effective after the numerical experiments of Miller et al. (1970) and Hahl (1970). 

It is remarkable that Lindblad developed the theory of density waves well before this mechanism was used in the waves of plasma, which has been very successful in later years.

B. Lindblad noticed that the theory of density waves was applicable both for trailing and leading waves. And as he was preoccupied with leading spiral arms, he applied the theory to them. But later the son of B. Lindblad made numerical experiments with a relatively large number of stars and he always found trailing spiral arms. Thus Prof. B. Lindblad devoted his  last two papers (1961, 1963) on trailing spiral arms.

The work of B. Lindblad on density wave spiral arms did not attract the attention it deserved because of his emphasis on leading spiral arms and its difficult style.

\begin{figure}[!h]
\centering
\includegraphics[scale=0.25]{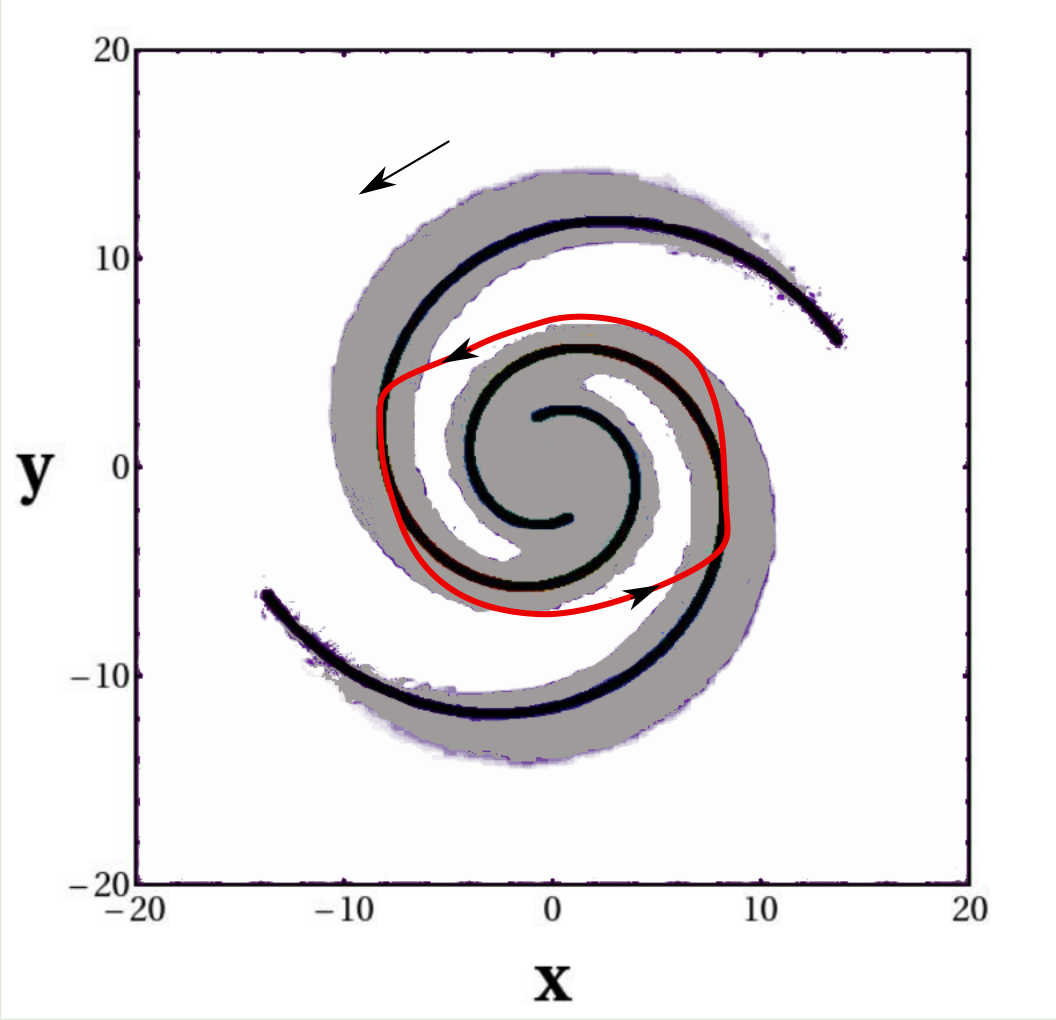}
\caption{The orbits in the density waves.}
\end{figure}

Then in 1963 I met Prof. C.C. Lin in MIT, who told me that he wanted to develop a theory of density waves to explain the spiral arms. I told him that such a theory was already developed by B. Lindblad. So, we went together to the library and borrowed some volumes of the Stockholms Observatoriums Annaler that contain the work of Lindblad. But when I met again Prof. Lin after a few days, he told me that he had difficulties to understand the papers of Lindblad. Thus, he decided to work the theory himself from scratch. In fact, one year later C.C. Lin and F. Shu published their first paper on the “grand design” density wave theory of spiral arms. A similar theory was developed by Kalnajs (1969, 1971).  

Their theory deals with the perturbations of the 3 basic functions of galactic dynamics, the potential V, the density $\rho$ and the distribution function $f$. These functions are related by 3 equations:

\begin{enumerate}

\item The collisionless Boltzmann equation

\begin{align}
\frac{df}{d{t}}=\frac{\partial{f}}{\partial{t}}+\bar{v}\frac{\partial f}{\partial\bar{x}}-\frac{\partial V}{\partial\bar{x}}\frac{\partial f}{\partial\bar{v}}=0,
\end{align}
where $\bar{x}=(x,y,z)$ and $\bar{v}=(v_x,v_y,v_z)$, 
that gives the distribution function $f$ (integral of motion) for a given potential $V$.

\item The integral
\begin{align}
\rho=\int f d \dot{\bar{x}},\quad d\dot{\bar{x}}=d\dot{x}d\dot{y}d\dot{z},
\end{align}
 that gives the density and 

\item the Poisson equation 
\begin{align}
\nabla^2V=4\pi G\rho
\end{align}
that connects the density with the potential.
\end{enumerate}

We consider a flat model $\rho=\sigma \delta(z)$, where $\sigma$ is the surface density, an axisymmetric background $V_0, f_0, \sigma_0$ and perturbations of first order $V_1,f_1,\sigma_1$ and higher orders $V=V_0+V_1+V_2+\dots$etc.

The perturbation $V_1$ of the potential is assumed to be of spiral form
\begin{equation}
V_1=A\exp[(i(\phi(r)+\omega t-m\theta],
\end{equation}
where $A$ is the amplitude and $m$ is the number of spiral arms (usually $m=2$ and $\omega=2\Omega$). 

The solution of the Boltzmann equation gives $f_1$. Then the integral (2) gives the ``response density'' $\sigma_1^{response}$ (linear theory). On the other hand the Poisson equation gives the imposed density $\sigma_1^{imposed}$ and we have to solve the self consistency equation
\begin{align}
\sigma_1^{imposed}=\sigma_1^{response}.
\end{align}

\begin{figure}[!h]
\centering
\includegraphics[scale=0.25]{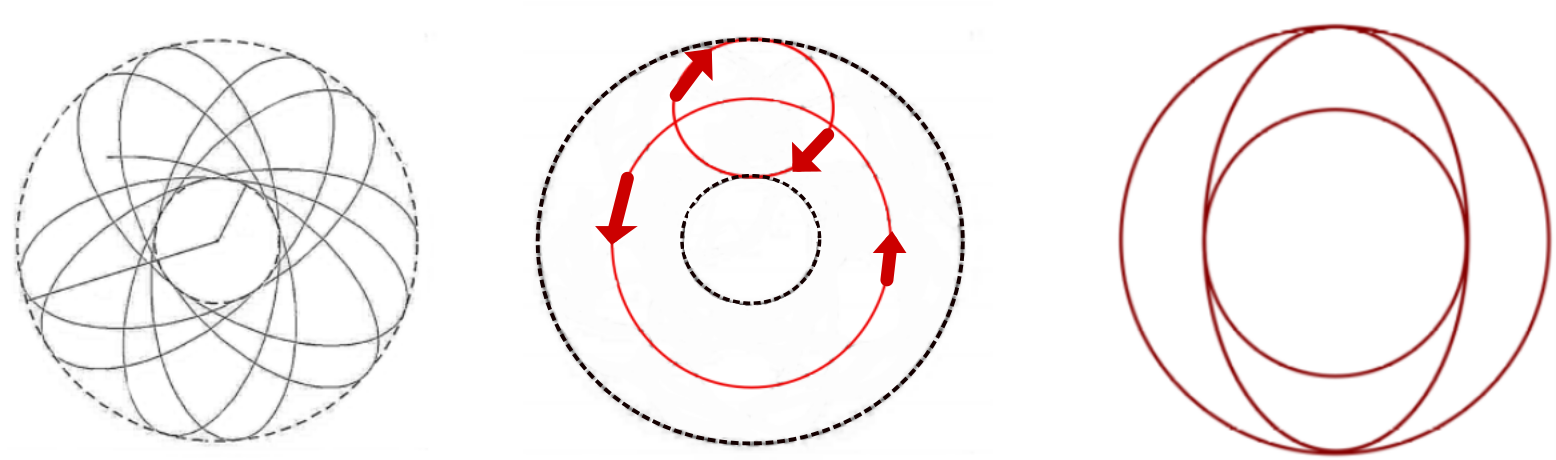}
\caption{Orbits in axisymmetric galaxies.}
\end{figure}

This is an integral equation (Kalnajs 1965, 1971). On the other hand  Lin and Shu (1964) found under some simplifying assumptions a  dispersion relation 
\begin{align}
1-\frac{2\pi G\sigma_0}{|k|\langle\dot{r}^2\rangle}\Big(1-\frac{v}{2\sin(\nu\pi)}\int_{-\pi}^{\pi}d\gamma\cos(\nu\gamma)\exp(-x(1+\cos(\gamma))\Big)=0,
\end{align}
where $\sigma_0$ is the unperturbed surface density and $\nu$ is the ratio of the frequencies
\begin{equation}
\nu=2(\Omega_s-\Omega)/\kappa,
\end{equation}
where $\Omega$ is the angular velocity of the stars at the distance $r$ and $\kappa$ is the corresponding epicyclic frequency.  Furthermore 
\begin{align}
\chi=\frac{k\langle \dot{r}^2\rangle}{\kappa^2}\end{align}
with $k$ the wavenumber ($k=\frac{2\pi}{\lambda}$, where $\lambda$ is the radial wavelength) and $\sqrt{\langle\dot{r}^2\rangle}$  the dispersion of the velocities.

The axisymmetric background ($V_0, f_0, \sigma_0$) is assumed to be given. The orbits in the background are rosettes (Fig. 4a) formed by combining a rotation around the center with an epicyclic motion (Fig. 4b). In the particular case of the Inner Lindblad Resonance the orbits are ellipses (Fig. 4c).

In Fig. 5 we have marked the values of $\Omega$  and of $\Omega-\kappa/2$ in two models. The pattern velocity of the spiral arms is $\Omega_s$. The ILR is when $\Omega-\kappa/2=\Omega_s$ (in which case $\nu=-1$).
In the first model (dotted lines) there is a single ILR, while in the second model (solid lines) there are two ILRs. The inner part of the spiral is between the outer ILR and corotation, where $\Omega=\Omega_s\, (\nu=0)$.

\begin{figure}[!h]
\centering
\includegraphics[scale=0.4]{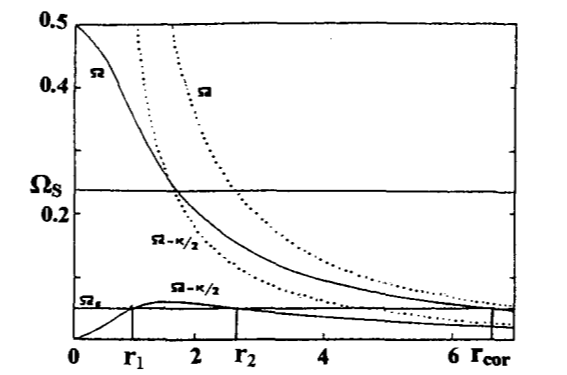}
\caption{The values of $\Omega$ and $\Omega-\kappa/2$ as functions of $r$ in two galactic models (dotted and solid lines). When $\Omega-\kappa/2=\Omega$ we have the Inner Lindblad Resonance.}
\end{figure}

The dispersion relation gives the wavelength of the spiral as a function of $\nu$ (which is a function of the distance $r$) (Fig. 6). The basic curve starts at $|\nu|=1$ (ILR) and terminates at $\nu=0$ (corotation). This represents a tight spiral wave between the ILR and  corotation. However there is also a second solution of an open spiral wave (large $\lambda$) that starts at corotation and reaches the ILR when $\lambda\to\infty$.

\begin{figure}[!h]
\centering
\includegraphics[scale=0.5]{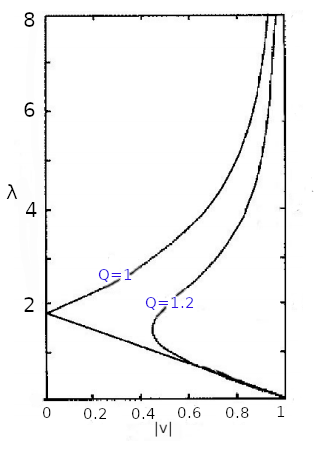}
\caption{ The Lin-Shu dispersion relation.}
\end{figure}

The two solutions of the dispersion relation are marked $Q=1$ (Fig.6). One tight solution that approaches asymptotically the ILR with zero wavelength  and  one open that approaches the ILR with very large $\lambda$. 

When $Q=1.2$ the two solutions are joined into one (Fig.6). The value of $Q$ is 
\begin{equation}
Q=\frac{\sqrt{\langle\dot{r}^2\rangle}}{\sqrt{\langle\dot{r}^2\rangle_{min}}},
\end{equation}
where $\sqrt{\langle\dot{r}^2\rangle_{min}}$ was given by Toomre (1964) and represents the minumum dispersion necessary to avoid  axisymmetric instabilities.
The spiral arms in a galactic model are given in Fig.7. These solutions are independent of the amplitude of the wave. The amplitude was given by Shu (1970).

\begin{figure}[!h]
\centering
\includegraphics[scale=0.2]{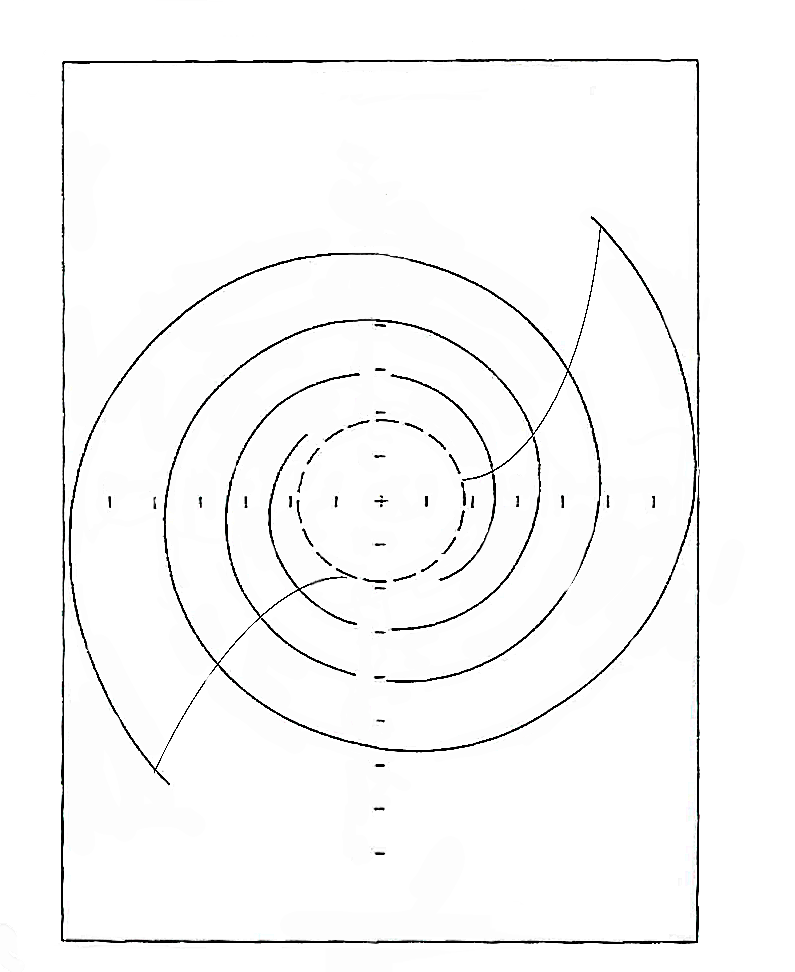}
\caption{The orbits in the original density wave theory.}
\end{figure}

After the work of Lin, Shu and Kalnajs, Toomre (1969) made an important contribution, by calculating the group velocity of the spiral density waves. The waves propagate with the group velocity inwards and after $10^9$ years they reach the Inner Lindblad Resonance where they are absorbed (Fig. 8). If so, the problem arises how the spiral arms are replenished and long lived. The solution of this problem by Toomre was a mechanism called “swing amplification”. Namely, the trailing spiral waves that move inwards reach the center of the galaxy and continue as open leading waves which become trailing later and reach corotation (Fig. 9).

\begin{figure}[!h]
\centering
\includegraphics[scale=0.4]{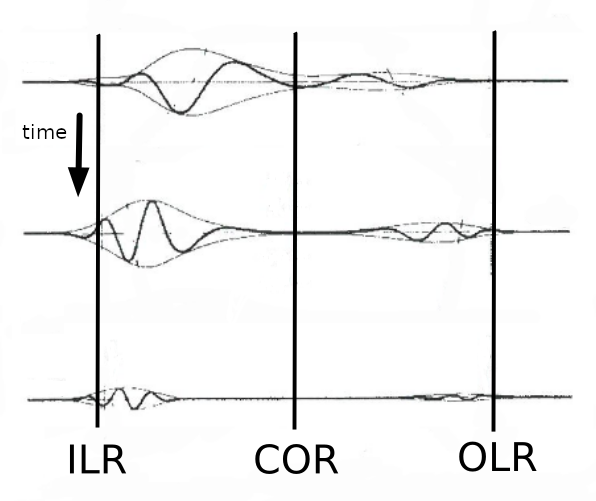}
\caption{The wave propagating inwards towards the ILR for $Q=1.2$.  A weaker wave propagates outwards to the OLR}
\end{figure}

\begin{figure}[!h]
\centering
\includegraphics[scale=0.4]{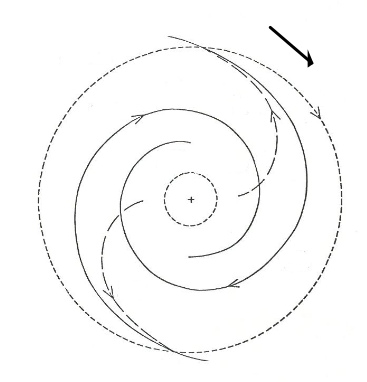}
\caption{The interaction of the various waves according to Lin and Mark. }
\end{figure}

 There they are enhanced by a local effect that was studied already by Julian and Toomre (1966) (Fig.~10) for stars and by Goldreich and Lynden-Bell (1965) for gas.  A local increase of density, like  the oval of Fig.~11  produces deviations of the surrounding orbits that increase further the density. Thus the wave grows (Fig.~9) but later it moves inwards with the group velocity (Fig.~9). Thus a cycle is formed that replenishes the spiral arms for long times.
 
\begin{figure}[!h]
\centering
\includegraphics[scale=0.5]{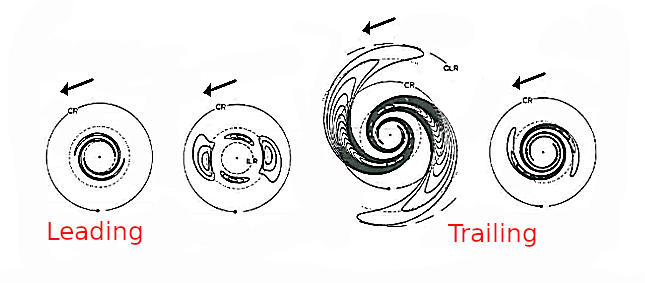}
\caption{The evolution of the initially leading spiral from the center of the galaxy that becomes           trailing as it approaches corotation and grows in amplitude. Then this wave moves inwards and decays.}
\end{figure}

A similar mechanism was proposed by Lin (1970) and Mark (1971, 1976) and was called “Waser”. Instead of a leading reflected wave they considered a trailing open wave, namely the second wave of Lin and Shu, as propagating outwards (Lin and Shu 1979) (Fig.~11). Further work on this topic was done by Lynden-Bell and Kalnajs (1972) and others.
 
A new advance in the theory of spiral density waves was made by considering nonlinear phenomena.

A preliminary form of a nonlinear density wave theory was already deve\-loped by Contopoulos and Woltjer in 1964. This paper was produced after a heated discussion between Woltjer and myself in Chicago (1963) in which Chandra\-sekhar was also involved. I had developed a theory giving a series solution of the Boltzmann equation $f=f_0+f_1+f_2+\dots$ in galactic dynamics (Contopoulos 1960) that was called a “third integral” (third after the energy and the angular momentum along the axis of symmetry). Woltjer argued that such a development, although valid for a smooth potential, would not be valid in a potential with small scale irregularities, like the spiral arms of a galaxy. However, I could convince Woltjer that the third integral should be applicable, if the amplitude of the spiral arms was sufficiently small. Thus, we decided to work on a rough nonlinear model of spiral arms and our paper was published next year (1964).

\section{Nonlinear effects at resonances}
The usual form of the integral $f=f_0+f_1+f_2+…$ is not applicable near the resonances. In particular, a drawback of the Lin-Shu dispersion relation was that it leads to infinities at the Lindbrad Resonances where $|\nu|=1$

The resonances is a well-known problem in the theory of the third integral. Namely the usual form of the third integral has several small divisors that become zero at various resonances.   In fact the terms of the third integral are of the form 
\begin{equation}
f_1=\frac{M_i}{m_1\omega_1-m_2\omega_2},
\end{equation}
where $\omega_1$ and $\omega_2$ are the basic frequencies and $m_1,m_2$ are integers. If $\omega_1/\omega_2=$ rational (resonance) there is a combination $|m_1\omega_1-m_2\omega_2|$ which is zero. If we are close to the resonance the quantity $|m_1\omega_1-m_2\omega_2|$ is small. This is called `small divisor'. To avoid this problem one has to write a new form of the third integral at every resonance (Contopoulos 1963, 1965) without the bad small divisor.

In the galactic case the new developments are necessary in particular at the ILR, the OLR and corotation. In fact in the dispersion relation there is a denominator $\sin(\nu\pi)$, where $\nu=2(\Omega_S-\Omega)/\kappa$ and this becomes $\mp 1$ and at the inner  and  outer Lindblad resonances this divisor becomes zero.

\begin{figure}[!h]
\centering
\includegraphics[scale=0.4]{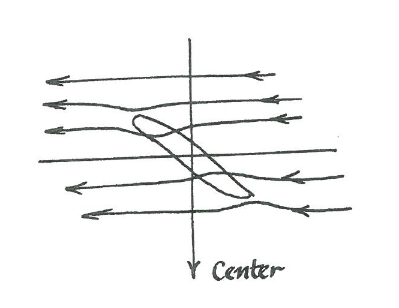}
\caption{Local enhancement of the density.}
\end{figure}

\begin{figure}[!h]
\centering
\includegraphics[scale=0.4]{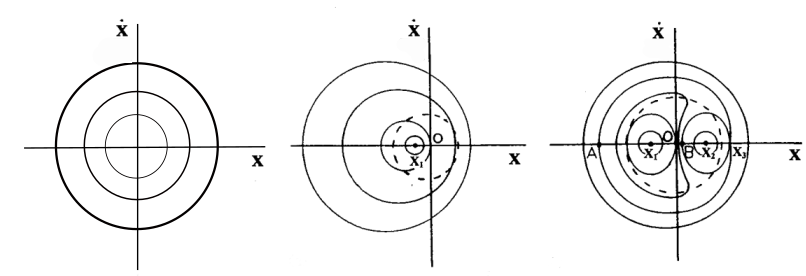}
\caption{The distribution of stars in the axisymmetric background.
b) in the spiral case away from the ILR c) near the inner Lindblad Resonance.}
\end{figure}

The dispersion relation does not become singular at corotation, where $\nu=0$, because then the factor $\frac{\nu}{\sin(\nu\pi)}$ becomes $1/\pi$. However, there is a divisor in the formula for the amplitude that goes to zero at corotation.

In the axisymmetric case the orbits are “rosettes” and their distribution is decided by equidensity curves which are like circles in the $(x,\dot{x})$ plane (Fig. 12a). In the spiral case away from the resonances the distribution is slightly different (Fig. 12b). However, near the ILR there are two populations of stars (Fig.12c) (Contopoulos 1975). The orbits of the two populations fill elliptical ovals perpendicular to each other (Fig. 13) (Contopoulos 1970). If we approach the ILR resonance from the inner side, we start with a distribution of type of Fig.~12a and then a small second distribution appears, which grows as we proceed outwards (Fig.12c). Beyond the resonance the first distribution shrinks and disappears and only the second distribution remains. Similar phenomena appear at the coro\-tation and at the Outer Lindblad Resonance. 

\begin{figure}[!h]
\centering
\includegraphics[scale=0.4]{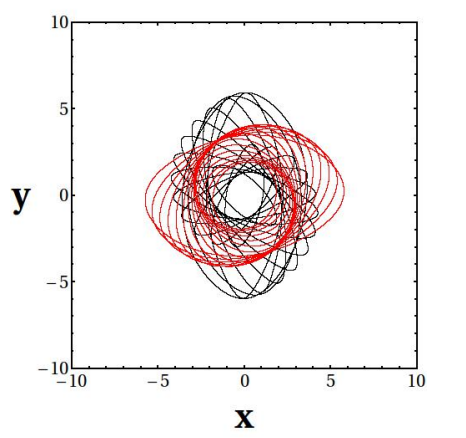}
\caption{The orbits near the ILR. }
\end{figure}

\begin{figure}[!h]
\centering
\includegraphics[scale=0.5]{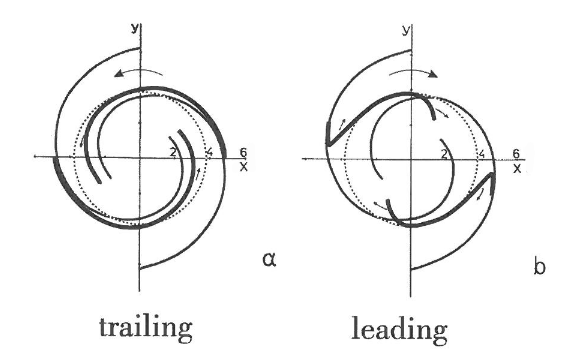}
\caption{Preference of trailing spiral arms. }
\end{figure}

The ILR plays also a role in supporting trailing rather than leading spiral waves (Contopoulos 1971). Namely, the spirals outside and inside the inner Lindblad resonance are joined by trailing arcs if the outer spiral is trailing (Fig. 14a). But if the outer spiral is leading then the joining  is trailing (Fig. 14b) and that changes the form of the wave from leading to trailing. 

\begin{figure}
\centering
\includegraphics[scale=0.35]{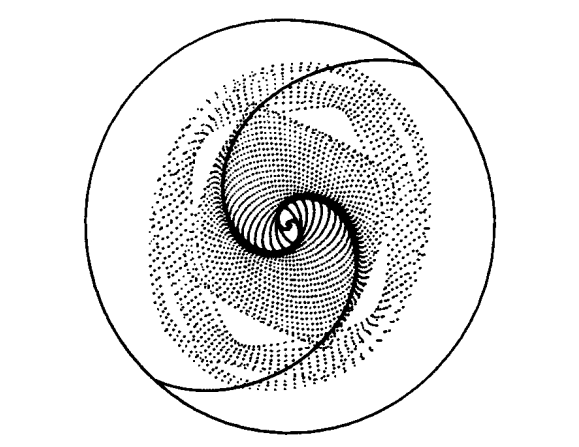}
\caption{Periodic orbits in normal galaxies. }
\end{figure}

Another characteristic of the spiral density waves refers to  the termination of the spirals. The usual density waves can be built by orbits close to the stable periodic orbits in the galaxy. 
In normal galaxies the periodic orbits in the main bodies of the galaxies are perturbed ellipses (Fig. 15) (Contopoulos 1985) with orientations that change gradually outwards. These orbits are called ``precessing ellipses'' (precessing not in time but in the radial initial conditions). The maxima of the density are along the spiral arms. However, when we approach the 4/1 resonance the orbits become almost rectangular (Fig. 15) and beyond the 4/1 resonance the rectangular orbits  change abruptly their orientation and do not support the continuation of the spiral arms. 
This is true for most normal spirals that terminate near the 4/1 resonance, if their amplitude is of order 5-10\% of the axisymmetric background. Only in spirals of order $<5\%$ one can have weak extensions up to corotation. An example of a galaxy whose spiral arms terminate near the 4/1 resonance is shown in Fig.~16.

\begin{figure}[!h]
\centering
\includegraphics[scale=0.5]{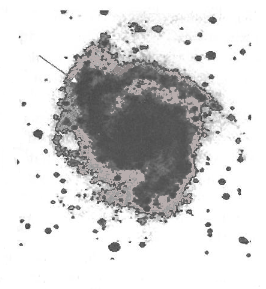}
\caption{Termination at the 4:1 resonance of the galaxy NGC 1997.}
\end{figure}

\section{Spirals in barred galaxies}

In barred galaxies the perturbations of the axisym\-metric background are very large of order 50-100\%. In this case most of the orbits inside corotation are elongated along the bar and support the bar.  But outside corotation most orbits are elongated perpendicularly to the bar (Fig.~ 17), therefore the bar terminates near corotation.

In the case of barred galaxies there are spirals beyond corotation. But in this case the orbits are mostly chaotic, as the perturbations are large.

\begin{figure}[!h]
\centering
\includegraphics[scale=0.22]{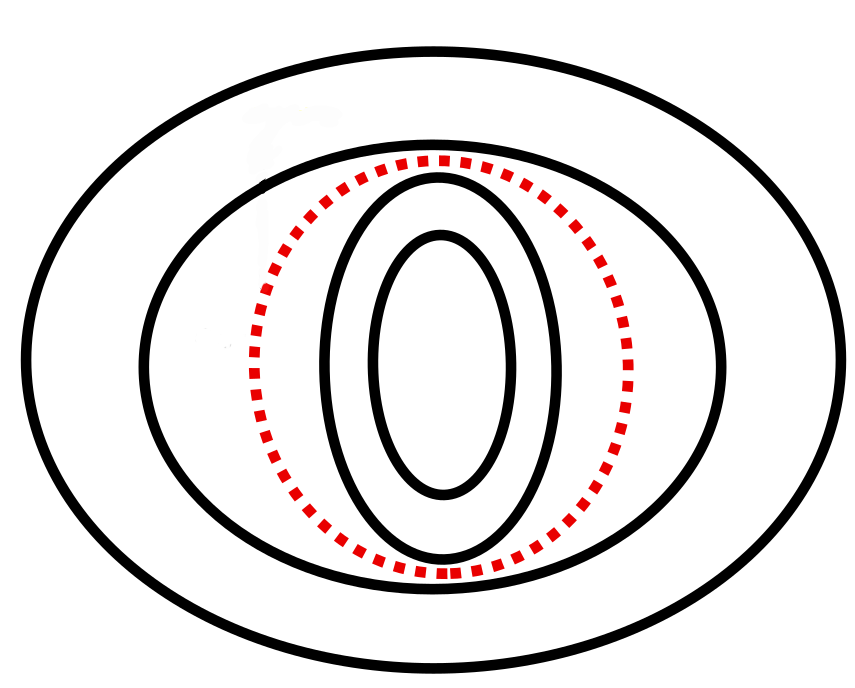}
\caption{The orbits in a barred galaxy support the bar only up to corotation (dotted circle).}
\end{figure}

Near the ends of the bar there are two unstable points, the Lagrangian points $L_1$ and $L_2$ (Fig.~ 18). Their unstable asymptotic curves are given as solid lines, and their stable curves as dashed lines. Twon unstable asymptotic curves form trailing spiral arms and the other two form an envelope of the bar.  The chaotic orbits have their apocenters along the two trailing spiral arms. Thus, although the orbits are chaotic, they form density waves with maximum density along the spiral arms.

 A first example of such an orbit was given by Kaufman and Contopoulos (1996) (Fig.~19). Later, Voglis et al. (2006) found the distribution of ordered and chaotic orbits in simulations of barred galaxies. The ordered orbits are in the main body of the bar, while the chaotic orbits populate the spirals and the outer envelope of the bar (Fig.~20). Similar results were found by Romero-Gomez et al. (2006).

\begin{figure}[!h]
\centering
\includegraphics[scale=0.4]{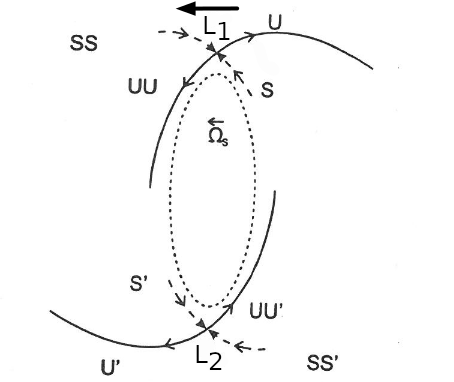}
\caption{ The Lagrangian points $L1, L2$ in a barred galaxy and its asymptotic curves.}
\end{figure}

\begin{figure}[!h]
\centering
\includegraphics[scale=0.3]{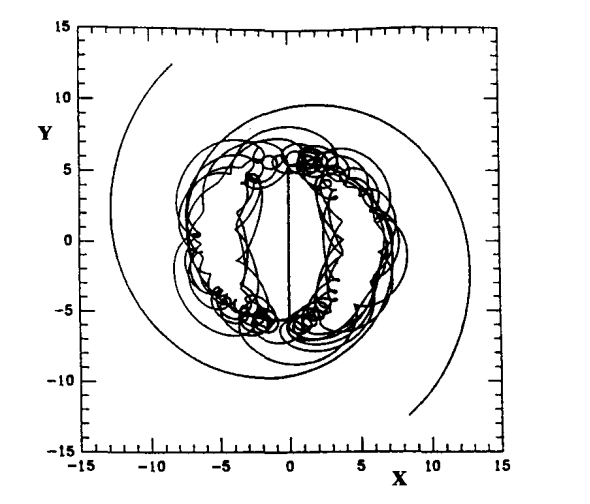}
\caption{A chaotic orbit in a barred galaxy. }
\end{figure}

\begin{figure}[!h]
\centering
\includegraphics[scale=0.4]{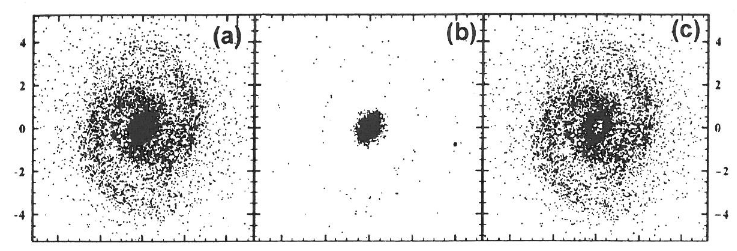}
\caption{The distribution of stars in N-body simulations of barred galaxies a) all the stars b) stars in ordered orbits c) stars in chaotic orbits.}
\end{figure}

These density waves are different from the usual density waves of the normal galaxies inside corotation where the orbits of the stars pass through the spiral arms but stay longer close to them (Fig.~3). In  normal galaxies most velocities are roughly perpendicular to the spiral arms. On the other hand, in the bar case the orbits are chaotic but their apocenters are close to the spiral arms and the average velocities are along these spiral arms (Fig.~21) (Patsis 2006).

\begin{figure}[!h]
\centering
\includegraphics[scale=0.4]{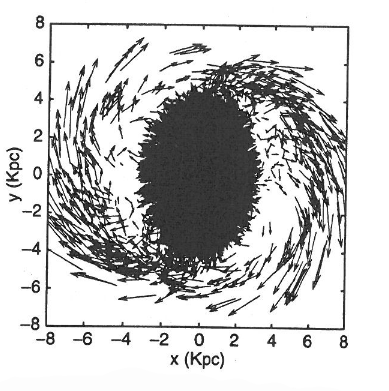}
\caption{The velocities of stars beyond corotation along the spiral arms. }
\end{figure}

\begin{figure}[!h]
\centering
\includegraphics[scale=0.3]{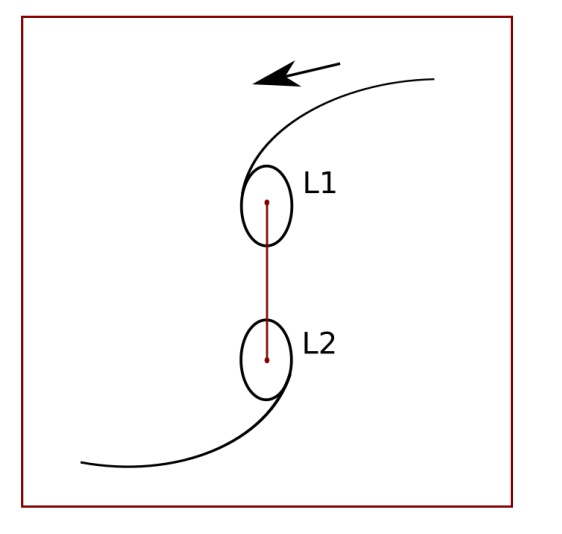}
\caption{ Periodic orbits near L1 and L2 in the case of 2 pattern speeds. }
\end{figure}

Some people have proposed models of barred galaxies that have different pattern velocities for the bar and for the spiral arms (Sellwood et al. 1988, 1989, 1991). In these cases, instead of the Lagrangian points $L_1$ and $L_2$ we have unstable periodic orbits (Fig. 23). The main asymptotic manifolds of these orbits are along the spiral arms. This is why these theories are called `manifold theories'. 
The other asymptotic manifolds are along the envelope of the bar. 

The theory of the manifolds of unstable periodic orbits has started with the work of Moser (1956, 1958) and Giorgilli (2001). Moser extended the theory of the third integral in cases near unstable periodic orbits. The usual form of the third integral is applied  to non-resonance cases near stable periodic orbits. This form does not have zero divisors, but it has many small divisors and, because of that, the series giving the integral does not converge. Nevertheless, if we truncate this integral at a particular high order we find very good approximate results (Contopoulos 2002). 

\begin{figure}[!h]
\centering
\includegraphics[scale=0.4]{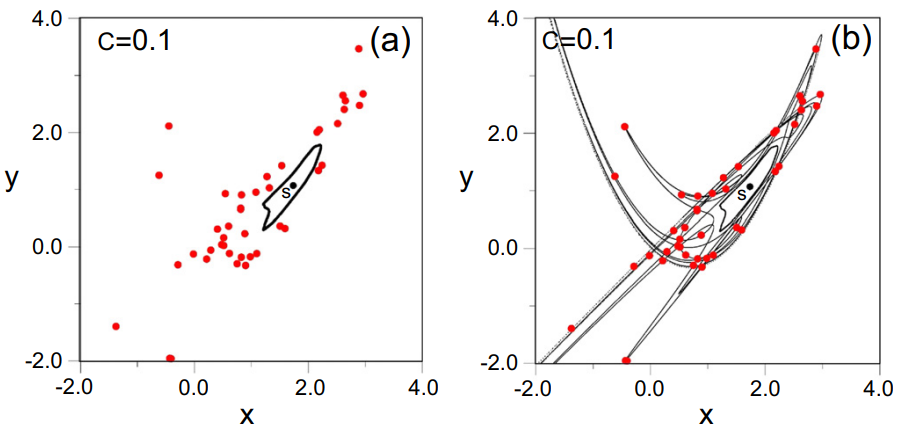}
\caption{The successive images of an initial point close to an unstable periodic orbit. }

\end{figure}

\begin{figure}[!h]
\centering
\includegraphics[scale=0.5]{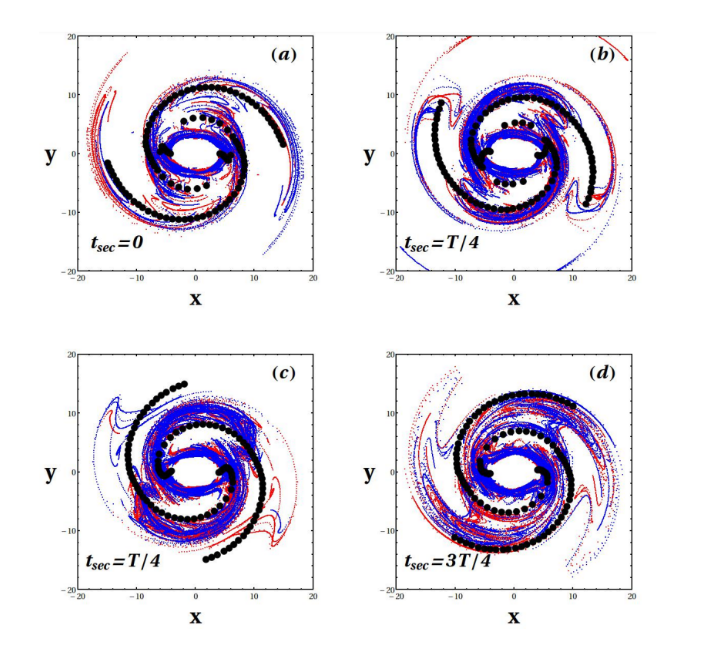}
\caption{The spiral arms in the case of 2 different pattern speeds of the bar and the spirals at different times. }
\end{figure}

On the other hand is the case of an unstable orbit one frequency ($\omega_1$) is real and the other ($\omega_2=i|\omega_2|$) is imaginary. Thus the divisors are of the form ($m_1\omega_1-im_2|\omega_2|)$ and they never approach zero. Because of that the third integral in this case is convergent. Then, although the orbits close to the unstable periodic orbit are chaotic, their positions at various times can be predicted with great accuracy (Efthymiopoulos, Contopoulos and Katsanikas 2014). As an example we see in Fig.~23 the successive intersections of an orbit with a surface of section starting on an  unstable manifold of the unstable periodic orbit (0,0) (Harsoula, Contopoulos and Efthymiopoulos 2015). These intersections (red points) are scattered in a chaotic way (Fig.~23a). However, all these points lie on the same asymptotic curve from the origin (Fig.~23b) and they can be given analytically with the help of convergent series.

This theory was applied by Efthymiopoulos, Harsoula and Contopoulos (2020) in different galactic problems with very good results. In particular, it was applied to the case of barred spiral galaxies with different pattern speeds of the bar and the spirals. 
The spirals found analytically describe very well the distribution of the orbits found by N-body simulations at different times (Fig.~24).

\section{Conclusions}

The main conclusion of our review is that there are two types of density wave spirals
\begin{enumerate}
\item Precessing ellipses in  normal galaxies inside corotation composed of ordered orbits.
\item Manifold spirals in barred galaxies. outside corotation composed of chaotic orbits.
\end{enumerate}
Thus the density wave theory has been successful both for normal galaxies and for barred galaxies. Its applications continue up to the present time, with the help of the new developments in Dynamical Astronomy.

\section*{Bibliography}
\begin{itemize}
\item Contopoulos, G.: 1960, Z. Astrophys. 49, 273.
\item Contopoulos, G.: 1963, Astron. J. 68, 763.
\item Contopoulos, G.: 1965, Astrophys. J. Suppl. 13, 503.
\item Contopoulos, G.: 1970, (in Becker, W. and Contopoulos, G. eds) IAU Symp. 38, 303.
\item Contopoulos, G.: 1971, Astrophys. J. 163, 181.
\item Contopoulos, G.: 1975, Astrophys. J. 201, 566.
\item Contopoulos, G.: 1985, Comments Astrophys. 11, 1.
\item Contopoulos, G.: 2002, Order and Chaos in Dynamical Astronomy, Springer Verlag. 
\item Contopoulos, G. and Patsis, P.A. (eds): 2009, Chaos in Astronomy. Springer Verlag. 
\item Contopoulos, G. and Woltjer, L.: 1964, Astrophys. J. 140, 1106.
\item Efthymiopoulos, C., Contopoulos, G. and Katsanikas, M.: 2014, Cel. Mech. Dyn. Astronom. 119, 321.
\item Efthymiopoulos, C., Harsoula, M. and Contopoulos, G.: 2020, Astron. Astrophys. 36, A44.
\item Giorgilli, A.: 2001, Discrete, Contin. Dyn. Syst. 25, 757.
\item Goldreich, P. and Lynden-Bell, D.: 1965, Month. Not. Roy. Astron. Soc. 130, 97.
\item Julian, W.H. and Toomre, A: 1966, Astrophys. J. 146, 810.
\item Harsoula, M., Contopoulos, G., and Efthymiopoulos, C.: 2015, J. Phys. A. 48, 135102.
\item Harsoula, M., Efthymiopoulos, C. and Contopoulos, G.: 2016, Month. Not. Roy. Astron. Soc. 459, 3419.
\item Harsoula, M., Zouloumi, K., Efthymiopoulos, C. and Contopoulos, G: 2021, Astron. Astrophys.  655, A55.
\item Hohl, F.: 1970, (in Becker, W. and Contopoulos, G. eds) IAU Symp. 38, 368.
\item Kalnajs, A.J.: 1965, PhD Thesis, Harvard Univ.
\item Kalnajs, A.J.: 1971, Astrophys. J. 166, 275.
\item Kaufmann D.E. and Contopoulos, G.: 1996, Astron. Astrophys.  309, 381.
\item Lin, C.C.: 1970, (in Becker, W. and Contopoulos, G.) IAU Symp. 38, 377.
\item Lin, C.C.: 1971, (in Jager, C. ed) Highlight of Astronomy 2, 88.
\item Lin, C.C.: 1976, (in Koiter, W.J. and Jager, C. eds) Theor. Appl. Math.  57a.
\item Lin, C.C. and Lau, Y.Y.: 1979, Stud. Appl. Math. 60, 97.
\item Lin, C.C. and Shu, F. H.: 1964, Astrophys. J. 140, 646.
\item Lin, C.C. and Shu, F. H.: 1966, Proc. Not. Roy. Astron. Soc. 55, 229.
\item Lindblad, B.: 1926, Upsala Med. No. 13. 
\item Lindblad, B.: 1936, Stockholm Obs. Ann. 12, No. 4. 
\item Lindblad, B.: 1940, Astrophys. J. 92, 1.
\item Lindblad, B.: 1941, Stockholm Obs. Ann. 13, No. 10. 
\item Lindblad, B.: 1961, Stockholm Obs. Ann. 21, No. 8. 
\item Lindblad, B.: 1963, Stockholm Obs. Ann. 32, No. 5. 
\item Lindblad, C.C. and Langebartel, H.: 1953, Stockholm Obs. Ann. 17, No.6.
\item Lynden-Bell, D. and Kalnajs, A.J.: 1972, Month. Not. Roy. Astron. Soc. 157, 1. 
\item Miller, R.H., Prendergait, K. H. and Quick, W.J.: 1970a, Astrophys. J. 61, 903.
\item Miller, R.H., Prendergait, K. H. and Quick, W.J.: 1970b, (in Becker, W. and Contopoulos, G., eds) IAU Symp. 38, 433. 
\item Moser, J.: 1956, Commun. Pure Appl. Math. 9, 673. 
\item Moser, J.: 1958, Commun. Pure Appl. Math. 11, 257. 
\item Patsis, P.: 2006, Month. Not. Roy. Astron. Soc. 369, L 56.
\item Romero-Gomez, M., Masdemont, J.J., Athanassoula, E. and Garcia-Gomez, C.: 2006, Astron. Astrophys. 453, 39.
\item Sellwood, J.A. and Sparke, L.S.: 1988, Astrophys. J. 425, 530.
\item Sellwood, J.A. and Lin, D.S.C.: 1989, Month. Not. Roy. Astron. Soc. 240, 991. 
\item Sellwood, J.A. and Kahn, F.D.: 1991, Month. Not. Roy. Astron. Soc. 250, 258. 
\item Shu, F.H.: 1968, PhD Thesis, Harvard Univ.
\item Shu, F.H.: 1970, Astrophys. J. 160, 89 and 99.
\item Toomre, A.: 1969, Astrophys. J., 158, 899.
\item Toomre, A.: 1977, Ann. Rev. Astron. Astrophys. 15, 437. 
\item Toomre, A.: 1981, (in Fall, S.M. and Lynden-Bell, D. eds) The structure and Dynamics of Normal Galaxies, Cambridge Univ. Pres. III.
\item Voglis, N., Stavropoulos, I. and Kalapotharakos, C.: 2006, Month. Not. Roy. Astron. Soc. 372, 901.
\end{itemize}

\end{document}